\documentclass{svproc}

\usepackage{url}
\usepackage{epsfig}  
\usepackage{graphicx}

\def\be{\begin{equation}}  
\def\ee{\end{equation}}  
 \def\Tr{{\rm Tr}\,}

\begin{document}
\mainmatter            
\title{Nuclear level densities: from empirical models to microscopic methods}
\titlerunning{Nuclear Level Densities}  
\author{Y. Alhassid}
\institute{Center for Theoretical Physics,  Sloane Physics Laboratory,  Yale University\\ New Haven, Connecticut 06520, USA\\
\email{yoram.alhassid@yale.edu},\\ WWW home page:
\texttt{https://alhassidgroup.yale.edu}}

\maketitle            
\begin{abstract}
The level density is among the most important statistical nuclear properties. It appears in Fermi's golden rule for transition rates and is an important input to the Hauser-Feshbach theory of compound nucleus reactions. We discuss empirical models of level densities and summarize the main experimental methods used to determine them. 
The microscopic calculation of level densities in the presence of correlations is a challenging many-body problem. We review recent microscopic approaches to calculate level densities. Mean-field and combinatorial methods have been applied across the nuclear chart, but often need to be augmented with empirical collective enhancement factors. The moment method and the auxiliary-field quantum Monte Carlo (AFMC) method are formulated in the context of the configuration-interaction shell model approach, and include correlations beyond the mean-field approximation.
\end{abstract}
\section{Introduction}
The nuclear level density is among the most important statistical nuclear properties. It appears in Fermi's golden rule for transition rates.  Along with gamma strength functions, it is a required input to the Hauser-Feshbach theory~\cite{Hauser1952} of compound nuclear reactions. The excited compound nucleus can decay into various channels, and its decay rate in any given channel is proportional to the available phase space, i.e., the corresponding level density of the residual nucleus.
The level density has many applications in diverse areas such as stellar nucleosynthesis and nuclear reactor technology.

The state density at total energy $E$ is defined as the number of states per unit energy
\be
\rho(E) = \Tr \delta(E-\hat H)\;,
\ee
where $\hat H$ is the system's Hamiltonian. For a system with discrete energy levels $E_i$, the state density $\rho(E) = \sum_i \delta(E-E_i)$ is singular.  Usually we are interested in a smoothed version of this density, i.e., the average state density.

While qualitative features of level densities can be understood by simple models, a quantitative understanding presents a major challenge, in
particular in the presence of correlations beyond the mean-field approximation.

The outline of this brief review is as follows. In Sec.~\ref{thermodynamics-approach} we discuss the thermodynamics approach for calculating level densities, which is based on calculation of the nuclear partition function at finite temperature.  In Sec.~\ref{Fermi-gas} we discuss the level density of non-interacting fermions, known as the Fermi gas level density, and simple models for the spin and parity distributions.  In Sec.~\ref{experimental-methods}  we summarize experimental methods used to measure level densities. In Sec.~\ref{empirical-models} we review the main empirical models for level densities, namely, the back-shifted Fermi gas model, the constant-temperature formula and the composite (Gilbert-Cameron) formula.  We then describe the major microscopic approaches for calculating level densities. In Sec.~\ref{mean-field} we discuss the mean-field approximation and the combinatorial method. Methods based on the configuration-interaction (CI) shell model that take into account correlations beyond the mean field are discussed in Secs.~\ref{moment-method} and \ref{AFMC}. In Sec.~\ref{moment-method} we discuss spectral averaging theory, which is based on the calculation of moments of the Hamiltonian. In Sec.~\ref{AFMC} we review the auxiliary-field quantum Monte Carlo (AFMC) method for calculating level densities and its applications. 

\section{Thermodynamics Approach}\label{thermodynamics-approach}

\subsection{Canonical ensemble}\label{canonical-ensemble}
We assume the nucleus to be in contact with a heat reservoir at temperature $T$, in which case its equilibrium configuration is described by the canonical Gibbs ensemble $e^{-\beta \hat H}$, where $\beta=1/T$ is the inverse temperature and $\hat H$ is the Hamiltonian. 

The partition function $Z(\beta) = \Tr\, e^{-\beta \hat H}$  is the Laplace transform of the state density $\rho(E)$, i.e., 
$Z(\beta) = \int_0^\infty dE e^{-\beta E} \rho(E)$. The level density is then the inverse Laplace transform of  the partition function
\be\label{inv-L} 
\rho(E) = {1 \over 2\pi i} \int_{-i \infty}^{i \infty} d\beta \,e^{\beta E} Z(\beta) \;. 
\ee
The inverse Laplace transform is numerically ill-defined. It can be evaluated
in the saddle-point approximation and provides the average level density~\cite{BM1969} 
\be\label{saddle-formula} 
\rho(E) \approx \left( 2\pi T^2 C \right)^{-1/2}  e^{S(E)}\;, 
\ee
where $S(E)$ is the canonical entropy and $C$ is the canonical heat capacity given by 
\be\label{S-C}
S =\ln Z + \beta E \;;\;\;\; C=\frac{dE}{dT} \;.
\ee
The value of $\beta$ used in Eqs.~(\ref{saddle-formula}) and (\ref{S-C})  is determined as a function of $E$ by the saddle-point condition 
\be\label{saddle-condition}
E = -{\partial \ln Z \over \partial \beta } = E(\beta) \;.
\ee

\subsection{Grand-canonical ensemble}\label{grand-canonical}

A similar thermodynamic approach can be followed in the grand-canonical ensemble, for which the number of particles fluctuates and only its average value is fixed.  The state density at energy $E$ and particle number $A$ is now given by a double inverse Laplace transform of the grand-canonical partition $Z_{\rm gc}(\beta,\alpha)= \Tr e^{-\beta H + \alpha \hat A}$ (the parameter $\alpha$ is related to the chemical potential $\mu$ by $\alpha=\beta \mu$).
In the saddle-point approximation we find~\cite{BM1969,Ericson1960}
 \be\label{saddle-state-GC}
\rho(E,A) \approx {1 \over 2\pi \sqrt{-\det D}} e^{S(E,A)} \;,
\ee
where $S=\ln Z_{\rm gc} +\beta E -\alpha A$ is the entropy, and $D$ is the $2\times 2$ matrix of second partial derivatives of $\ln Z_{\rm gc}$ with respect to $\beta$ and $\alpha$.
The values of $\beta$ and $\alpha$ are determined as a function of $E$ and $A$ from the saddle-point equations
\begin{eqnarray} \label{saddle-GC}
 -{\partial \ln Z_{\rm gc} \over \partial \beta}  =  E \;,\,\,\, {\partial \ln Z_{\rm gc} \over \partial \alpha}  =  A \;.
 \end{eqnarray}

\section{Non-interacting (Fermi Gas) Models}\label{Fermi-gas}

For non-interacting fermions, it is easier to use the grand-canonical formalism of Sec.~\ref{grand-canonical}. 

We first consider one type of nucleon. The logarithm of the many-particle grand-canonical partition function for non-interacting fermions is
\be
\ln Z_{\rm gc} =\int_0^\infty d \varepsilon g(\varepsilon) \ln \left[1 + e^{-\beta(\varepsilon -\mu)}\right] \;,
\ee
where $g(\varepsilon)$ is the single-particle density of states. 

The thermal energy can be calculated as a function of temperature using the low-temperature expansion of Sommerfeld~\cite{Sommerfeld1928} for temperature $T\ll T_F$ (where $T_F$ is the Fermi temperature).  To second order in $T$
\be\label{energy}
E=E_0 + a T^2 \;,
 \ee
 where $E_0$ is the ground-state energy and $a=\frac{\pi^2}{6} g (\varepsilon_F)$ ($\varepsilon_F$ is the Fermi energy, i.e., the energy of highest occupied single-particle level).
 
 The corresponding heat capacity is $C=dE/dT =2 a T$.  Using $C= T d S/dT$, we determine the entropy to be $S=2aT = 2 \sqrt{ a E_x}$, where  $E_x=E-E_0$ is the excitation energy.  The saddle-point approximation (\ref{saddle-state-GC}) then leads to Bethe's formula for one type of nucleon~\cite{Bethe1936} 
 \be
 \rho(E_x) = \frac{1}{\sqrt{48} E_x} e^{2 \sqrt{ a E_x}}  \;.
 \ee
 
 A similar derivation for both protons and neutrons with $Z\approx N$ gives~\cite{BM1969}
 \be\label{Bethe}
 \rho(E_x) = \frac{\sqrt{\pi}}{12} a^{-1/4} E_x^{-5/4} e^{2 \sqrt{a E_x}}\;,
 \ee
 where $a=\frac{\pi^2}{6}[ g_p(\epsilon^{(p)}_F) + g_n(\epsilon^{(n)}_F)]$.
For $Z \neq N$, the state density is given by an equation similar to Eq.~(\ref{Bethe}) but contains an additional factor of $g/(2\sqrt{g_p\, g_n})$ on its r.h.s.~(which is of order unity).

In the free Fermi gas model, assuming $A$ nucleons in a box, $a={\pi^2 A\over 4 \,\epsilon_F} \approx A / 15\;{\rm MeV}^{-1}$. 
A more realistic estimate is obtained for an isotropic harmonic oscillator potential, for which $a \approx A / 10\;{\rm MeV}^{-1}$.
Using a Woods-Saxon potential, it was found that $a\approx A/10.7$ MeV$^{-1}$ in medium-mass nuclei~\cite{Alhassid2003}.

\subsection{Spin-cutoff model}

The spin-cutoff model assumes random coupling of single-particle spins~\cite{Bethe1937,Ericson1960}. In this model, the distribution of the spin projection $M=\sum_i m_i$ is Gaussian
\be\label{M-dist}
\frac{\rho_M}{\rho} = \frac{1}{\sqrt{2\pi} \sigma}  e^{-M^2/2\sigma^2}\;,
\ee
where $\sigma$ is the spin-cutoff parameter.  Using the equipartition theorem at temperature $T$, we find
\be\label{sigma-I}
\sigma^2 = \frac{I T}{\hbar^2}\;,
\ee
with $I$ being the thermal moment of inertia. At higher excitation energies, $I$ approaches its rigid-body value~\cite{BM1969}, but it decreases at low excitation energies because of pairing correlations. 

The spin distribution is calculated from
\be
\rho_J =\rho_{M=J} - \rho_{M=J+1} \approx -\frac{d \rho_M}{dM} \Big |_{M=J+1/2} \;.
\ee
Using Eq.~(\ref{M-dist}), we find for the spin-cutoff model 
\be\label{J-dist} 
\frac{\rho_J}{\rho} = \frac{2J+1}{2\sqrt{2\pi} \sigma^3}  e^{-J(J+1)/2\sigma^2}\;.
\ee

\subsection{Parity distribution}

A simple model for the parity distribution of level densities is obtained by assuming the particles occupy the single-particle states independently and randomly~\cite{Ericson1960}. We divide the single-particle levels into two groups of positive and negative parities, and denote by $\pi$ the parity of the group with the  smaller occupation probability $p_\pi$. The probability to have $n$ particles in this group is then a binomial distribution
\be\label{binomial}
P(n) = {A \choose n} p_\pi^n (1-p_\pi)^{A - n}  \;,
\ee
where $A$ is the total number of excited particles.  For an even-particle system, a negative (positive) parity many-particle state corresponds to odd (even) values of $n$, and the total probability to have an negative (positive) parity is obtained by summing $P(n)$ over all odd (even) values of $n$. For small $p_\pi$ and large $A$, we can approximate (\ref{binomial}) by a Poisson distribution $P(n) = \frac{f^n}{n!} e^{-f_\pi} $, which depends on a single parameter $f=A p_\pi$, the total occupation of the $\pi$-parity orbitals.  For an even-particle system, the ratio of negative- to positive-parity partition functions at a given temperature is then given by~\cite{Alhassid2000}
\be\label{parity-ratio}
{Z_- \over Z_+} = \sum_{n \;{\rm odd}} P(n) / \sum_{n \;{\rm even}} P(n) =\tanh f \;.
\ee
Eq.~(\ref{parity-ratio}) holds more generally for an even-even nucleus with $f=f_p +f_n$ being the total average occupation of the $\pi$-parity orbitals for both protons and neutrons. For an even-even nucleus, the positive-parity states dominate at low excitations, but equilibration of both parities is achieved above a certain excitation energy. In practical applications, it is often assumed that the parity distribution is already equilibrated at the neutron resonance energy. 

\section{Experimental Methods}\label{experimental-methods}
The measurement of level densities is a challenging task. There are several methods but all have systematic uncertainties and are limited to certain energy regimes:

\begin{itemize}

\item  Level counting at low excitation energies. This requires the knowledge of a complete set of measured energy levels~\cite{ENSDF}. 

\item Neutron and proton resonance data~\cite{RIPL-3} provide an estimate of the level density at the neutron or proton threshold energy.  The measured resonance level spacing (usually $s$ wave and sometimes also $p$ wave) provides the level density at certain values of the spin/parity determined by the selection rules. The conversion to total densities requires a model for the spin distribution, and often a spin-cutoff model with rigid-body moment of inertia is used.  

\item Particle evaporation spectra~\cite{Voinov2019}, which depend on the level density through the Hauser-Feshbach formalism~\cite{Hauser1952}. This method requires the knowledge of particle transmission coefficients, which can be calculated from optical potential models.

\item  The ``Oslo method" which uses the measured particle and $\gamma$-ray coincidence matrix~\cite{Oslo}.  The extraction of level densities in this method requires the knowledge of level counting data at low energies and neutron resonance data. 

\end{itemize} 

Progress has often been achieved by combining several of these methods. 

\section{Empirical Models}\label{empirical-models}
Several phenomenological models have been introduced to describe level densities in the presence of correlations. 

\subsection{Back-shifted Fermi gas formula} 

Pairing correlations and shell effects are empirically taken into account in Bethe's formula by shifting the ground-state energy by a backshift parameter $\Delta$
 \be\label{BBF}
 \rho(E_x) = \frac{\sqrt{\pi}}{12} a^{-1/4} (E_x-\Delta)^{-5/4} e^{2 \sqrt{a (E_x - \Delta)}}\;.
 \ee
 This back-shifted Bethe formula for the state density includes two parameters $a$ and $\Delta$ that can be treated as adjustable parameters.  They can for example be determined from level counting data at low excitation energies and neutron resonance data~\cite{Dilg1973,vonEgidy1988}.
 
In the state density, each level with spin $J$ is counted $2J+1$ times (i.e., the magnetic degeneracy is included).  The level density is defined by counting only once each level with spin $J$. Assuming a spin-cutoff model (\ref{J-dist}), the level density $\tilde \rho$ is related to the state density $\rho$ by
\be\label{level-density}
\tilde \rho(E_x) = \sum_J \rho(E_x,J) = { \rho(E_x) \over \sqrt{2\pi}\sigma} \;.
\ee
 Global fits using an energy-dependent parameter $a$ that includes shell effects were carried out in Ref.~\cite{Koning2008}. 
 
 \subsection{Constant-temperature formula}
 
 At low excitation energies it is found empirically that the level density $\tilde \rho$ is well described by an exponential function 
 \be\label{constant-T}
 \tilde \rho (E_x)= \frac{1}{T_1} e^{(E_x-E_1)/T_1}
 \ee
 where $E_1$ and $T_1$ are parameters.  $T_1$ can be interpreted as an effective temperature
 \be
 T_1^{-1} = d \ln \tilde\rho(E_x)/d E_x \;.
\ee

\subsection{Composite (Gilbert-Cameron) formula}

The composite formula for the level density, also known as Gilbert-Cameron formula~\cite{Gilbert1965}, is a constant-temperature formula (\ref{constant-T}) at low energies and a back-shifted Fermi gas formula (\ref{level-density}) and (\ref{BBF}) at higher excitations. Both the level density and its first derivative are matched at a certain excitation energy $E_M$, so overall the composite formula has only two adjustable parameters.

\section{Mean-Field and Combinatorial Methods}\label{mean-field}

\subsection{Mean-field methods}

Hartree-Fock (HF) mean-field theory using Skyrme interactions plus finite-temperature BCS has been applied in Ref.~\cite{Demetriou2001} to the large number of nuclei that are involved in nucleosynthesis. 
 
A mean-field theory provides the intrinsic level density $\rho_{\rm int}(E_x)$. It has to be augmented by collective enhancement factors (vibrational and rotational)
 \be
 \rho(E_x) = K_{\rm vib}(E_x) K_{\rm rot}(E_x) \rho_{\rm int}(E_x) \;,
 \ee
 where the factors $K_{\rm vib}(E_x)$ and $ K_{\rm rot}(E_x)$ describe the enhancement of the density due to vibrational and rotational collective states.  
 The energy dependence of these factors, and in particular, their decay with excitation energy $E_x$,  is one of the least understood issues in studies of level densities, and are usually parametrized by phenomenological expressions~\cite{Capote2009}.
 
\subsection{Combinatorial methods}

\begin{figure}
\centerline{\includegraphics[width=\textwidth]{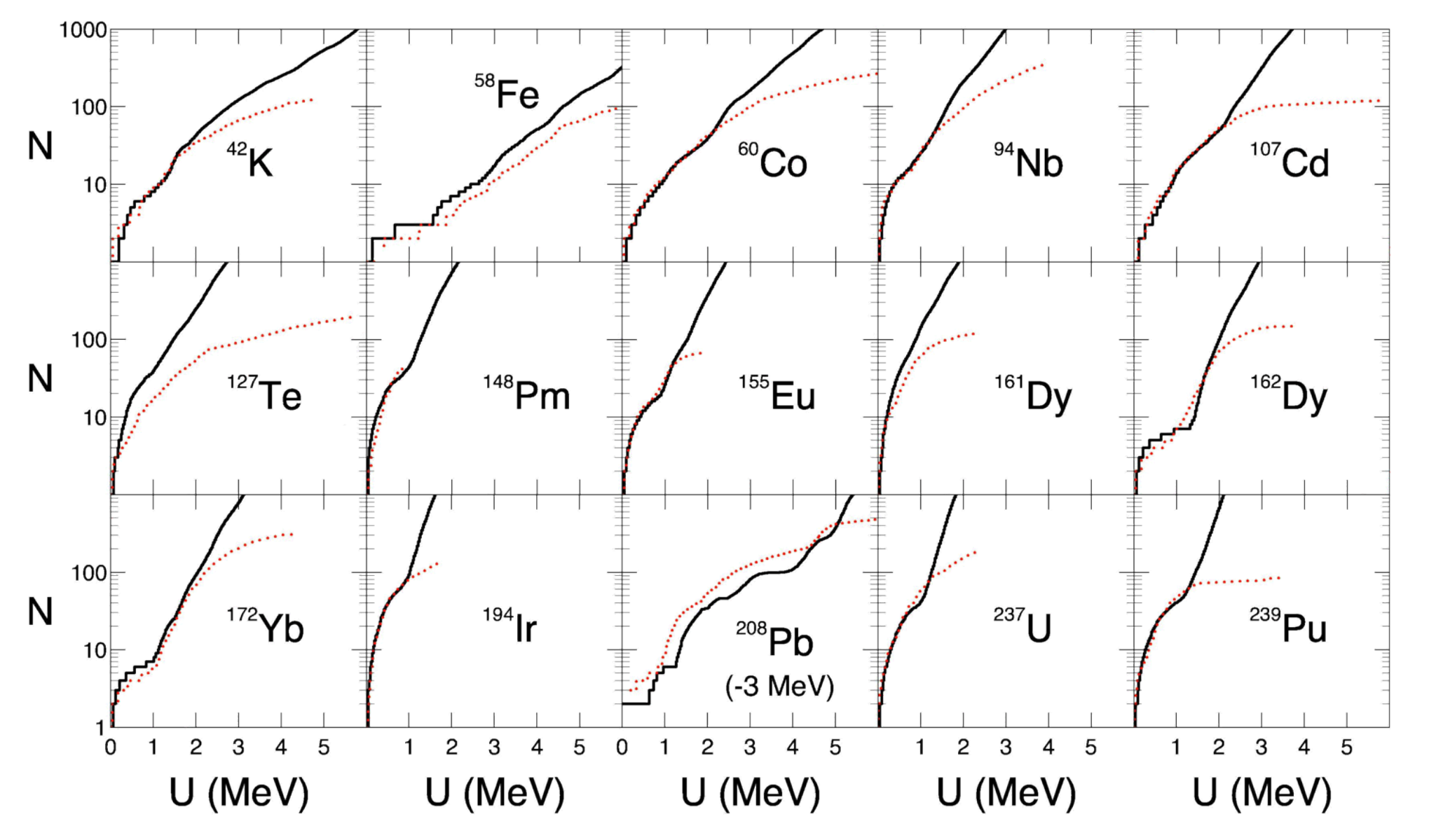}}
\caption{Cumulative level densities calculated in the combinatorial approach (solid histograms) are compared with cumulative number of observed levels (dotted histograms) at low excitation energy $U$.  Adapted from Ref.~\cite{Goriely2008}.}
\label{LD_combinatorial}
\end{figure}

The combinatorial models are based on counting the number of ways to distribute the nucleons among single-particle
levels at a given total excitation energy~\cite{Hilaire2006,Goriely2008,Hilaire2012,Uhrenholt2013}.  They are often combined with a mean-field theory such as the Hartree-Fock-Bogoliubov (HFB) approximation. Examples of cumulative level densities calculated in the mean field plus combinatorial approach are shown in Fig.~\ref{LD_combinatorial}.
 
\section{Configuration-interaction Shell Model Methods}\label{shell-model}
The CI shell model includes shell effects and correlations beyond the mean-field approximation, and thus can in principle provide the most precise microscopic calculation of level densities. However, the combinatorial growth of the dimensionality of the many-particle model space with the number of valence nucleons and/or the number of valence orbitals has hindered its application in mid-mass and heavy nuclei.

\subsection{Spectral Averaging Theory (Moment Method)}\label{moment-method}
The spectral averaging theory, also known as the moment method, describes the density as a superposition of Gaussian densities for various partitions of the single-particle orbitals  with centroids and widths that are determined by the first two moments of the Hamiltonian~\cite{Mon1975,Kota2010,Horoi2004,Senkov2013}.

The method requires a reliable calculation of the ground-state energy, which is required for determining the excitation energy. 
The calculation of second moments is time consuming in large model spaces, and so far the method has been applied to light and mid-mass nuclei, where it provides good agreement with experimental data and with exact CI shell model calculations (in $sd$-shell nuclei)~\cite{Senkov2016}.  For more details of the method and its applications see Refs.~\cite{Zelevinsky2019,Zelevinsky2019a}.

\subsection{Auxiliary-Field Quantum Monte Carlo Method}\label{AFMC}
The auxiliary-field quantum Monte Carlo (AFMC) method, also known in nuclear physics as the shell model Monte Carlo (SMMC)~\cite{Lang1993,Alhassid1994,Koonin1997,Alhassid2001,Alhassid2017}, is based on the Hubbard-Stratonovich (HS) transformation~\cite{HS-trans}, in which Gibbs ensemble $e^{-\beta \hat H}$  is written as a superposition of ensembles  $\hat U_\sigma$ describing non-interacting nucleons moving in external auxiliary fields $\sigma(\tau)$ 
\be \label{HS}
 e^{-\beta \hat H} = \int {\cal D}[\sigma] G_\sigma \hat U_\sigma \;,
\ee
where $G_\sigma$ is a Gaussian weight.  The calculation of the integrand for a given configuration of the auxiliary fields $\sigma$ reduces to matrix algebra in the single-particle space of typical dimension $\sim 50 -100$. The integration over the large number of auxiliary fields is carried out using Monte Carlo methods. 

The AFMC state density is calculated using the thermodynamic approach of Sec.~\ref{canonical-ensemble}~\cite{Nakada1997,Ormand1997,Langanke1998}. The canonical thermal energy $E(\beta)$ is calculated as a function of $\beta$ and Eq.~(\ref{saddle-condition}) is integrated to find the partition function $Z(\beta)$. The entropy and heat capacity are calculated from Eqs.~(\ref{S-C}), and the average state density is then given by Eq.~(\ref{saddle-formula}).

\subsubsection{Mid-mass nuclei}

 AFMC methods were applied to mid-mass nuclei using the complete $fpg_{9/2}$ shell~\cite{Nakada1997,Alhassid1999,Alhassid2007,Mukherjee2012}. The single-particle levels and orbitals are taken from a Woods-Saxon potential with spin-orbit interaction.  The two-body interaction includes the dominating components~\cite{Dufour1996} of effective nuclear interactions: monopole pairing and multipole-multipole interactions with quadrupole, octupole and hexadecupole components.  

\begin{figure}[h]
\centerline{\includegraphics[width=0.8\textwidth]{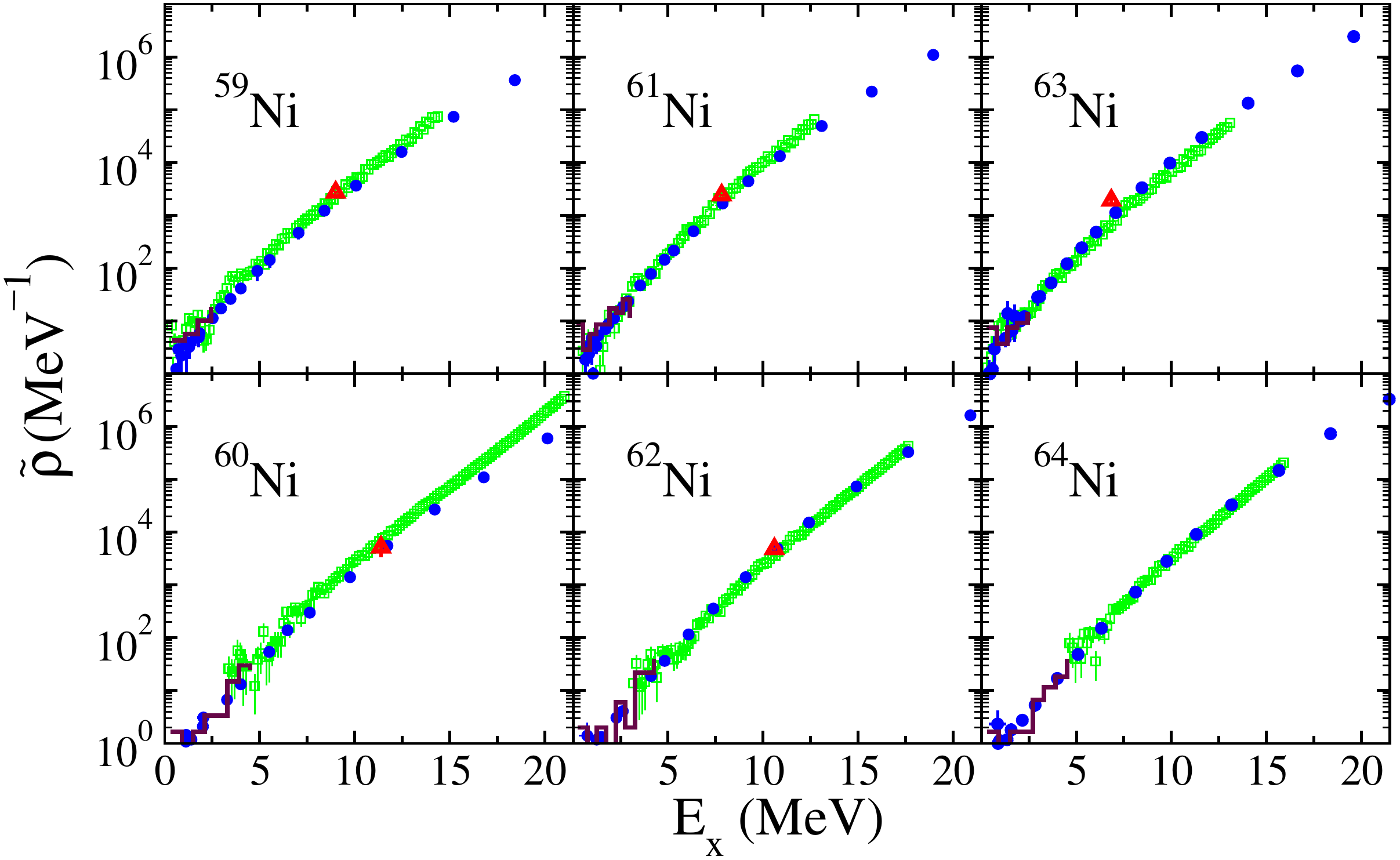}}
\caption{Level densities of $^{59-64}$Ni isotopes versus excitation energy $E_x$.  The AFMC level densities (blue circles) are compared with level densities determined by proton evaporation experiments (green symbols)~\cite{Voinov2012}, neutron resonance data when available (red triangles), and level counting data at low excitation energies (blue histograms). Taken from Ref.~\cite{Bonett2013}.}
\label{nickel}
\end{figure}

 AFMC level densities of nickel isotopes $^{59-64}$Ni are shown by the blue circles in Fig.~\ref{nickel}~\cite{Bonett2013} . These densities do not include the magnetic degeneracy $2J+1$ of each level with spin $J$ and are obtained by projection on $M=0$ for even-mass nuclei and $M=1/2$ for odd-mass nuclei~\cite{Alhassid2015}.  The AFMC densities are in excellent agreement with experimental data without any adjustable parameters.

\subsubsection{Heavy nuclei: the lanthanides}

The AFMC approach was extended to the proton-neutron formalism, in which protons and neutrons can occupy different shells~\cite{Alhassid2008}.  This formulation was used to study chains of samarium and neodymium isotopes which exhibit a crossover from vibrational to rotational collectivity as a function of the number of neutrons.  The corresponding CI shell model space includes the complete $50-82$ shell plus $1f_{7/2}$ orbital for protons, and the complete $82-126$ shell plus the $0h_{11/2}$ and $1g_{9/2}$ orbitals for neutrons. 

\begin{figure}[h!]
\centerline{\includegraphics[width=\textwidth]{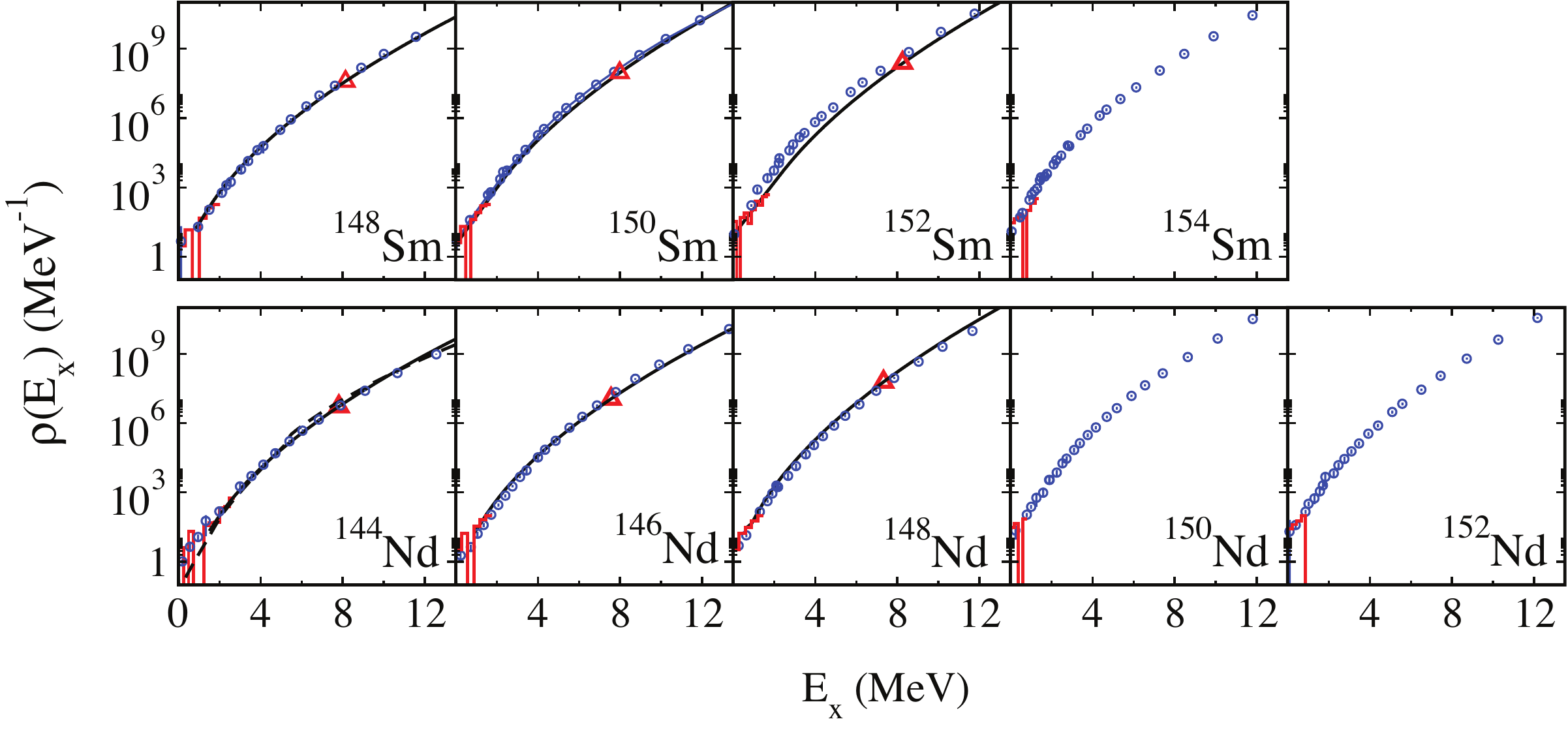}}
\caption{State densities in even-mass samarium and neodymium isotopes vs.~excitation energy $E_x$.  The AFMC  densities (blue circles) are compared with level counting data (histograms) at low excitation energies, and with neutron resonance data (triangles) when available.  
 Adapted from Refs.~\cite{Ozen2013,Alhassid2014a}.}
 \label{LD_lanthanides}
\end{figure}

Fig.~\ref{LD_lanthanides}  shows AFMC state densities (open circles) for chains of samarium and neodymium isotopes~\cite{Ozen2013,Alhassid2014a}. Good agreement is seen with experimental data.

\subsubsection{Rotational enhancement in deformed nuclei}

 Finite-temperature mean-field approximations to level densities were benchmarked in Ref.~\cite{Alhassid2016} against exact AFMC results. The mean-field approximation is formulated in the grand-canonical ensemble, and it is necessary to project on fixed number of protons and neutrons to compare with the canonical AFMC results.  Particle-number projection was carried out using various approximations (including the saddle-point approximation) and by exact projection after variation~\cite{Fanto2017}. 

\begin{figure}[b!]
\centerline{\includegraphics[width=0.75\textwidth]{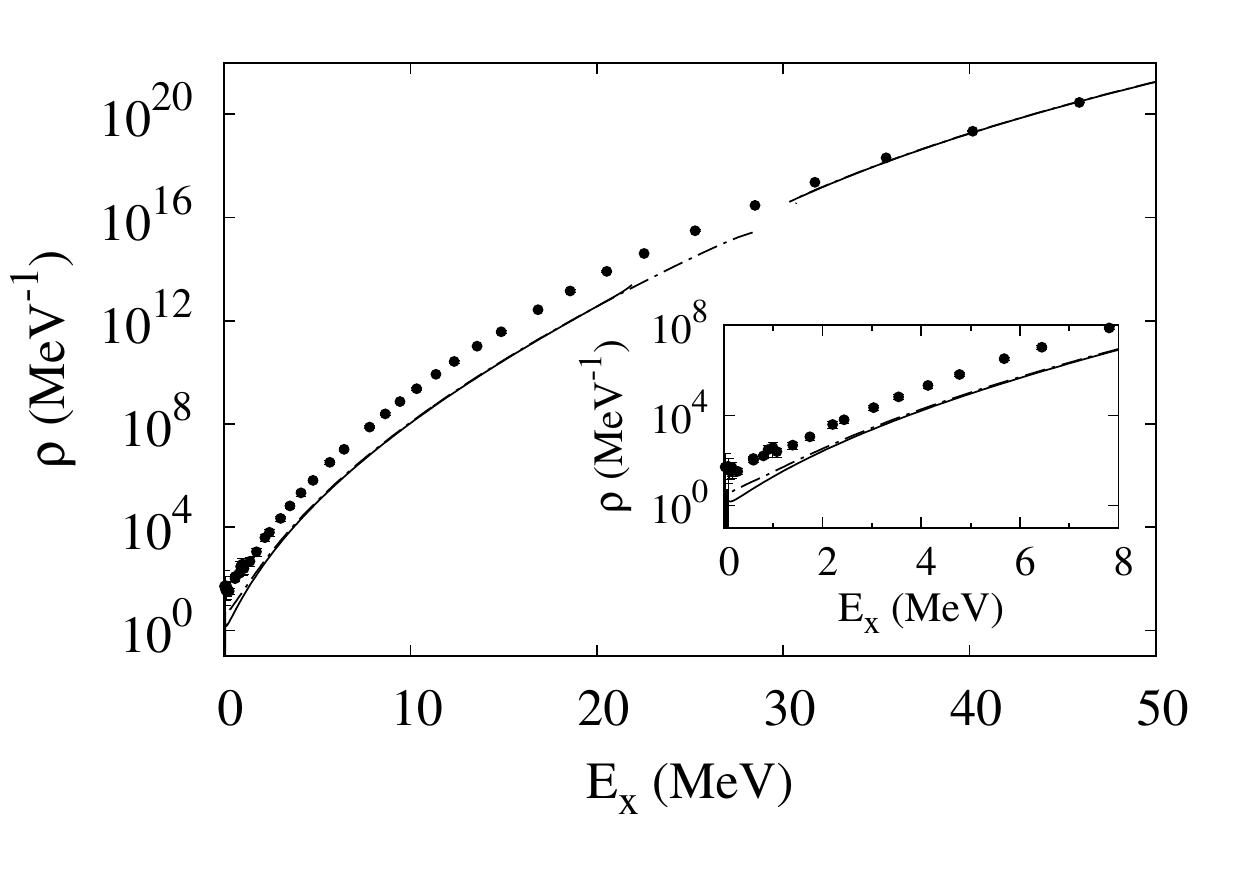}}
\caption{State density of $^{162}$Dy vs.~excitation energy $E_x$. The AFMC state density (solid circles) is compared with the HF density (solid line). The inset shows the low excitation energy region. Taken from Ref.~\cite{Alhassid2016}.}
\label{Dy162}
\end{figure}

In Fig.~\ref{Dy162},  the mean-field HF level density of a deformed nucleus $^{162}$Dy  is compared with the AFMC density.  The HF describes the intrinsic states, and thus the enhancement of the exact AFMC density (compared with HF density) is due to rotational bands that are built on top of the intrinsic bandheads. The corresponding rotational enhancement factor decays to 1 in the vicinity of the mean-field shape transition ($E_x \sim 30$ MeV) from a deformed to a spherical shape.

\subsubsection{Spin and parity distributions}  Exact spin projection was implemented in AFMC and used to calculate the spin distributions in mid-mass nuclei~\cite{Alhassid2007}. It was found that the spin-cutoff model works well except at low excitation energies in even-even nuclei for which a staggering effect in spin was observed. 

\begin{figure}
\centerline{\includegraphics[width=0.9\textwidth]{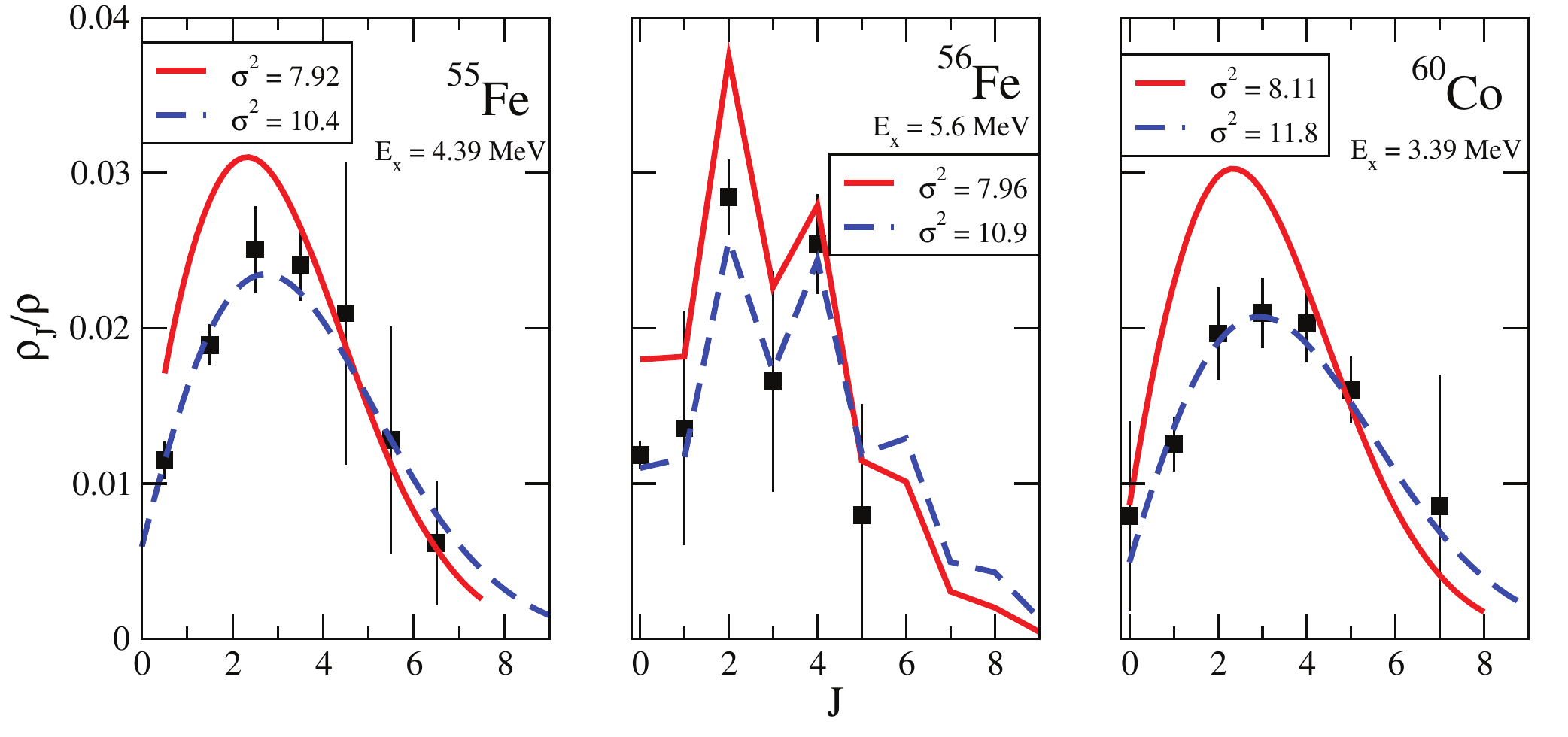}}
\caption{Spin distributions $\rho_J/\rho$ versus $J$ for $^{55}$Fe, $^{56}$Fe and $^{60}$Co.  The solid squares are the AFMC results of Ref.~\cite{Alhassid2007}, and the solid lines are empirical distributions~\cite{vonEgidy2008,vonEgidy2009}. The dashed lines are obtained from the solid lines by scaling the spin-cutoff parameter $\sigma$ to larger values, taking into account the larger excitation energies used in the AFMC calculations.  Taken from Ref.~\cite{vonEgidy2008}.}
\label{spin-dist}
\end{figure}

Fig.~\ref{spin-dist} shows spin distributions $\rho_J/\rho$ as a function of spin $J$ for the odd-even nucleus $^{55}$Fe, the even-even nucleus $^{56}$Fe, and the odd-odd nucleus $^{60}$Co.  AFMC results are compared with empirical distributions determined from the analysis of complete sets of experimentally known nuclear energy levels~\cite{vonEgidy2008,vonEgidy2009}.  A staggering effect in spin can be seen in $^{56}$Fe.

The spin-cutoff parameter can be related to the thermal moment of inertia through Eq.~(\ref{sigma-I}). For even-even nuclei, the moment of inertia is found to be suppressed below the pairing transition~\cite{Alhassid2007,Alhassid2015}.

Exact parity projection was also implemented in AFMC~\cite{Nakada1997,Alhassid2000}.  The resulting parity distributions in mid-mass nuclei were found to be well described by Eq.~(\ref{parity-ratio}) when, below the pairing transition temperature, $f$ is taken to be the average occupation of the quasi-particle states with parity $\pi$.
 
\subsubsection{The deformation dependence of level densities}

Modeling of shape dynamics, e.g., fission, requires knowledge of the level density as a function of intrinsic deformation. The theory of deformation has mostly relied on mean-field approximation that breaks rotational invariance.

 In Ref.~\cite{Mustonen2018}  a model-independent method was developed to calculate distributions of intrinsic deformation within the rotationally invariant framework of the CI shell model without invoking a mean-field approximation.  The method uses a projection on the axial quadrupole operator in the laboratory frame~\cite{Alhassid2014,Gilbreth2018}, and is based on a Landau-like expansion of the logarithm of the quadrupole shape distribution in quadrupole invariants~\cite{Kumar1972,Cline1986} up to fourth order. We note that this expansion is similar to the Landau expansion of the free energy used to describe shape transitions in nuclei with the quadrupole deformation playing the role of the order parameter~\cite{Levit1984,Alhassid1986}.
 
\begin{figure}[bth]
\begin{minipage}{0.6\textwidth}
 \includegraphics[width=0.6\textwidth]{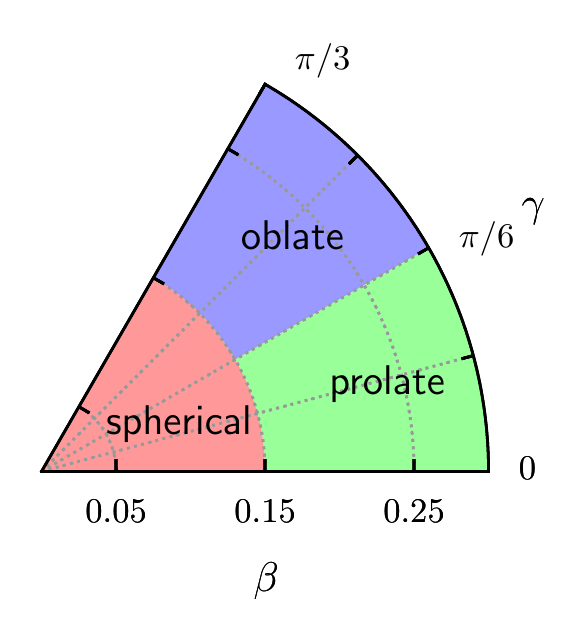}
 \end{minipage}
\begin{minipage}{0.4\textwidth} \caption{Three shape regions in the $\beta-\gamma$ plane of intrinsic quadrupole deformation.}
 \label{shapes}
\end{minipage}
\end{figure}

\begin{figure}[]
\centerline{\includegraphics[width=1.1\textwidth]{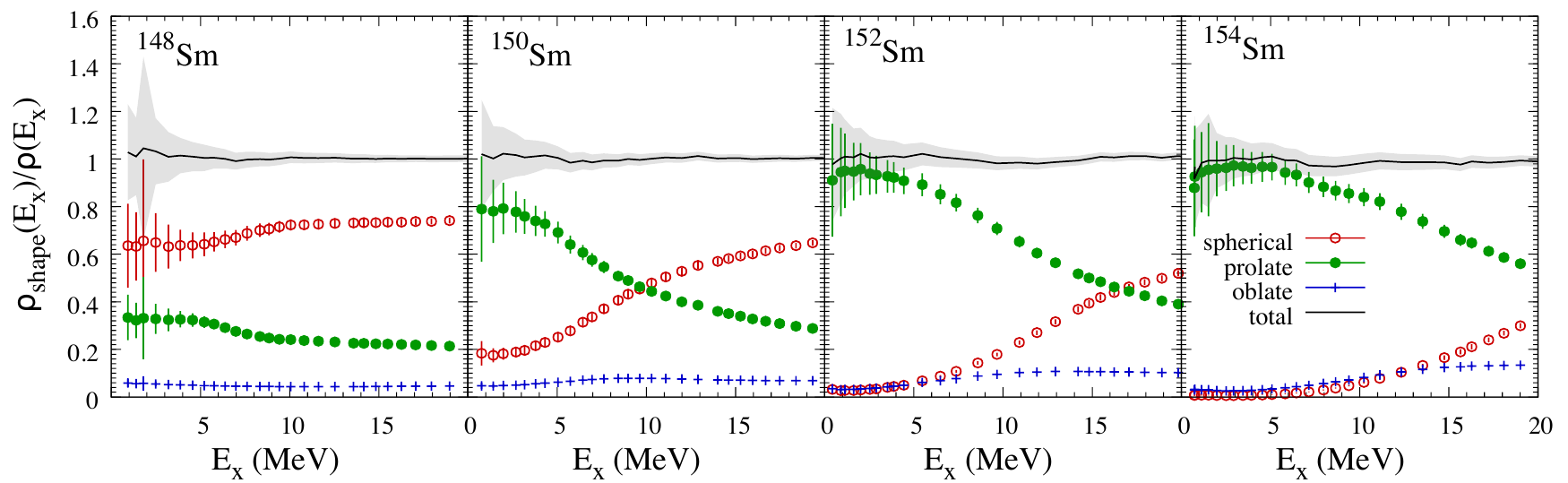}}
\caption{Fraction $\rho_{\rm shape}(E_x)/\rho(E_x)$ of the state density in each of the three intrinsic deformation regions of Fig.~\ref{shapes} for even-mass $^{148-154}$Sm isotopes vs.~excitation energy $E_x$. Taken from Ref.~\cite{Mustonen2018}.}
\label{fraction-shape}
\end{figure}

 The method of Ref.~\cite{Mustonen2018} enables the calculation of shape-dependent state densities $\rho(E_x,\beta,\gamma)$ as a function of excitation energy $E_x$ and intrinsic quadupole deformation parameters $\beta,\gamma$. To facilitate the presentation of the shape-dependent densities, the $\beta-\gamma$ plane is divided into three regions: spherical, prolate and oblate as shown in Fig.~\ref{shapes}, and $\rho(E_x,\beta,\gamma)$ is integrated over each one of these regions using the metric $4\pi^2\beta^4 |\sin 3\gamma | \, d\beta \, d\gamma$ to obtain $\rho_{\rm shape}(E_x)$. In Fig.~\ref{fraction-shape}, the fraction $\rho_{\rm shape}(E_x)/\rho(E_x)$ of the state density in each of these three regions is shown as a function of excitation energy for spherical ($^{148}$Sm), transitional ($^{150}$Sm) and deformed ($^{152}$Sm,$^{154}$Sm) nuclei.    
 
 As is seen in Fig.~\ref{fraction-shape}, the spherical region dominates the state density in  $^{148}$Sm.  In the deformed $^{152}$Sm and  $^{154}$Sm nuclei, the prolate region dominates the state density at lower excitation energies but the spherical density becomes comparable and exceeds the prolate density at higher excitations where a shape transition occurs in the mean-field approximation.

\section{Conclusion}\label{conclusion}

Phenomenological models of level densities are often based on empirical
modifications of the Fermi gas model and on the constant-temperature formula.

 Mean-field and combinatorial models are the most common microscopic approaches to level densities and have been applied across the table of nuclei. However, they must be supplemented by empirical collective enhancement factors.
 
 The moment method and the auxiliary-field Monte Carlo (AFMC) method 
include correlations beyond the mean-field approximation within the framework of the CI shell model.  The moment method has been applied to light and mid-mass nuclei, while AFMC has been applied to nuclei as heavy as the lanthanides.\\

\noindent {\bf Acknowledgments.}  I would like to thank G.F. Bertsch, M. Bonett-Matiz, L. Fang, P. Fanto, C.N. Gilbreth, S. Liu,  A. Mukherjee, M.T. Mustonen, H. Nakada, and C. \"Ozen for their collaboration on parts of the work reviewed above. This work was supported in part by the U.S. DOE grants Nos.~DE-SC0019521 and DE-FG02-91ER40608. 
The research presented here used resources of the National Energy Research Scientific Computing Center, which is supported by the Office of Science of the U.S. Department of Energy under Contract No.~DE-AC02-05CH11231.  We also thank the Yale Center for Research Computing for guidance and for use of the research computing infrastructure.

% ---- Bibliography ----
%


\begin{thebibliography}{99}

\bibitem{Hauser1952} W.~Hauser and H.~Feshbach, Phys. Rev. {\bf 87}, 366 (1952).
\bibitem{BM1969} A. Bohr and B. R. Mottelson, {\em Nuclear Structure},  Vol.~1 (Benjamin, New York, 1969).
\bibitem{Ericson1960} T. Ericson, Adv. Phys. {\bf 9}, 425 (1960).
\bibitem{Sommerfeld1928} A. Sommerfeld, Zeitschrift Physik {\bf 47}, 1 (1928). 
\bibitem{Bethe1936} H. A. Bethe, Phys. Rev. {\bf 50}, 332 (1936).
\bibitem{Alhassid2003} Y. Alhassid, G.F. Bertsch, and L. Fang, Phys. Rev. C {\bf 68}, 044322 (2003).
\bibitem{Bethe1937} H. A. Bethe, Rev. Mod Phys. {\bf 9}, 69 (1937).
\bibitem{Alhassid2000} Y. Alhassid, G.F. Bertsch, S. Liu, and H. Nakada, Phys. Rev. Lett. {\bf 84}, 4313 (2000).
\bibitem{ENSDF}  For discrete level scheme data, see the ENSDF database at \texttt{https://www.nndc.bnl.gov/ensdf/}.
\bibitem{RIPL-3} For neutron and proton resonance data, see the RIPL-3 database at \texttt{https://www-nds.iaea.org/RIPL-3/}.
\bibitem{Voinov2019} A.V. Voinov, this proceedings.
\bibitem{Oslo} A. Schiller, L. Bergholt, M. Guttormsen, E. Melby, J. Rekstad, and S. Siem,  Nucl. Instrum. Meth. Phys. Res. A {\bf 447}, 498 (2000).
\bibitem{Dilg1973}  W. Dilg, W. Schantl,H. Vonach and M Uhl, Nucl. Phys. A {\bf 217}, 269 (1973). 
\bibitem{vonEgidy1988} T. von Egidy, H. Schmidt, and A. Bekhami, Nucl. Phys. A {\bf 481}, 189 (1988). 
\bibitem{Koning2008} A. Koning, S. Hilaire, and S. Goriely, Nucl. Phys. A {\bf 810}, 13 (2008).
\bibitem{Gilbert1965} A. Gilbert and A.G.W. Cameron, Can. J. Phys. {\bf 43}, 1446 (1965).
\bibitem{Demetriou2001} P. Demetriou and S. Goriely, Nucl. Phys.A {\bf 695}, 95 (2001).
\bibitem{Capote2009} R. Capote et al.,
 Nuclear Data Sheets {\bf 110},  3107 (2009).
 \bibitem{Hilaire2006}  S.~Hilaire and S.~Goriely, Nucl. Phys. A {\bf 779}, 63  (2006). 
\bibitem{Goriely2008} S. Goriely, S. Hilaire, and A. J. Koning, Phys. Rev. C {\bf 78}, 064307 (2008).
\bibitem{Hilaire2012} S. Hilaire, M. Girod, S. Goriely, and A. J. Koning, Phys. Rev. C {\bf 86}, 064317 (2012).
\bibitem{Uhrenholt2013} H. Uhrenholt, S.  {\AA}berg, A. Dobrowolski, T. D{\o}ssing, T. Ichikawa, and P. Moller,  Nucl. Phys. A {\bf 913}, 127 (2013).
\bibitem{Mon1975} K. Mon and J. French, Annals of Physics {\bf 95}, 90 (1975).
\bibitem{Kota2010}  {\em  Spectral Distributions in Nuclei and Statistical Spectroscopy},  edited by V.K.B. Kota and R.U. Haq (World Scientific, Singapore, 2010). 
\bibitem{Horoi2004} M. Horoi, M. Ghita, and V. Zelevinsky, Phys. Rev. C {\bf 69}, 041307 (2004).
\bibitem{Senkov2013} R. A. Sen'kov, M. Horoi, and V. Zelevinsky, Com. Phys. Comm. {\bf 184}, 215 (2013).
\bibitem{Senkov2016} R.A. Sen'kov and V. Zelevinsky, Phys. Rev. C {\bf 93}, 064304 (2016).
\bibitem{Zelevinsky2019} V. Zelevinsky and M. Horoi, Prog.  Part. Nucl. Phys. {\bf 105}, 180 (2019).
\bibitem{Zelevinsky2019a} V. Zelevinsky and S. Karampagia, this proceedings. 
\bibitem{Lang1993} G.H.~Lang,  C.W.~Johnson, S.E.~Koonin, and W.E.~Ormand, Phys. Rev. C {\bf 48}, 1518 (1993).
\bibitem{Alhassid1994} Y. Alhassid, D. J. Dean, S. E. Koonin, G. Lang, and W.E. Ormand, Phys. Rev. Lett. {\bf 72}, 613 (1994).
\bibitem{Koonin1997} S.E. Koonin, D.J. Dean, and K. Langanke, Phys. Rep. {\bf 278}, 2 (1997).
\bibitem{Alhassid2001} Y.~Alhassid, Int. J. Mod. Phys. B {\bf 15}, 1447 (2001).
\bibitem{Alhassid2017}  Y. Alhassid, in {\em Emergent Phenomena in Atomic Nuclei from Large-Scale Modeling: a Symmetry-Guided Perspective}, edited by K. D. Launey (World Scientific, Singapore, 2017) pp. 267 -- 298.
\bibitem{HS-trans} J. Hubbard, Phys. Rev. Lett. {\bf 3}, 77 (1959); R.L. Stratonovich, Dokl. Akad. Nauk. S.S.S.R. {\bf 115}, 1097 (1957).
\bibitem{Nakada1997} H. Nakada and  Y. Alhassid, Phys. Rev. Lett.~{\bf 79}, 2939 (1997).
\bibitem{Ormand1997} W.E. Ormand, Phys. Rev. C {\bf 56}, 1678 (R) (1997).
\bibitem{Langanke1998} K. Langanke, Phys. Lett. B {\bf 438}, 235 (1998).
\bibitem{Alhassid1999} Y. Alhassid, S. Liu, and H. Nakada, Phys. Rev. Lett.~{\bf 83}, 4265 (1999).
\bibitem{Alhassid2007} Y. Alhassid, S. Liu and H. Nakada, Phys. Rev. Lett. {\bf 99}, 162504 (2007).
\bibitem{Mukherjee2012}  A. Mukherjee and Y. Alhassid, Phys. Rev. Lett.~{\bf 109}, 032503 (2012).
\bibitem{Dufour1996}  M. Dufour and  A.P. Zuker, Phys. Rev. C {\bf  54}, 1641 (1996).
\bibitem{Bonett2013} M. Bonett-Matiz, A. Mukherjee, and Y. Alhassid, Phys. Rev. C  {\bf 88}, 011302 (R) (2013).
\bibitem{Alhassid2015} Y. Alhassid, M. Bonett-Matiz, S. Liu, and H. Nakada,  Phys. Rev. C {\bf 92}, 024307 (2015).
\bibitem{Voinov2012} A.V. Voinov, S.M. Grimes, C.R. Brune, T. Massey, and A. Schiller, EPJ Web of Conferences {\bf 21}, 05002 (2012);  A.V. Voinov (private communication).
\bibitem{Alhassid2008} Y. Alhassid, L. Fang, and H. Nakada, Phys. Rev. Lett. {\bf 101}, 082501 (2008).
\bibitem{Ozen2013} C. \"Ozen, Y. Alhassid, and H. Nakada, Phys. Rev. Lett.  {\bf 110}, 042502   (2013).
\bibitem{Alhassid2014a}  Y. Alhassid Y,  C. \"Ozen, and H. Nakada, Nuclear Data Sheets {\bf 118}, 233 (2014).
\bibitem{Alhassid2016}   Y. Alhassid, G.F. Bertsch, C.N. Gilbreth, and H. Nakada, Phys. Rev. C {\bf 93}, 044320 (2016). 
\bibitem{Fanto2017} P. Fanto, Y. Alhassid, and G.F. Bertsch, Phys. Rev. C {\bf 96}, 014305 (2017).
\bibitem{vonEgidy2008} T. von Egidy and D. Bucurescu, Phys. Rev. C {\bf 78}, 051301 (R) (2008). 
\bibitem{vonEgidy2009}T. von Egidy and D. Bucurescu, Phys. Rev. C {\bf 80}, 054310 (2009).
\bibitem{Mustonen2018} M.T. Mustonen, C.N. Gilbreth, Y. Alhassid, and G.F. Bertsch, Phys. Rev. C {\bf 98}, 034317 (2018).
\bibitem{Alhassid2014} Y. Alhassid, C.N. Gilbreth, and G.F. Bertsch, Phys. Rev. Lett.  {\bf 113}, 262503 (2014).
\bibitem{Gilbreth2018} C.N. Gilbreth, Y. Alhassid, and G.F. Bertsch, Phys. Rev. C {\bf 97}, 014315 (2018).
\bibitem{Kumar1972} K. Kumar, Phys. Rev. Lett. {\bf 28}, 249 (1972).
\bibitem{Cline1986} D. Cline, Ann. Rev. Nucl. Part. Sci. {\bf 36}, 683 (1986).
\bibitem{Levit1984} S. Levit and Y. Alhassid, Nucl. Phys. A {\bf 413}, 439 (1984).
\bibitem{Alhassid1986} Y. Alhassid, S. Levit, and J. Zingman, Phys. Rev. Lett. {\bf 57}, 539 (1986).

\end{thebibliography}
\end{document}